\documentclass[prd,onecolumn,preprintnumbers,nofootinbib]{revtex4}

\usepackage{graphicx}

\usepackage[plainpages=false, colorlinks=true, anchorcolor=blue, linkcolor=blue, citecolor=blue, bookmarks=false]{hyperref}
\usepackage{color}
\usepackage{float}
\usepackage{amsfonts,amsmath,amssymb}
\usepackage{natbib}
\usepackage{array}
\usepackage{subcaption}
\usepackage{caption}
\usepackage{booktabs}
\newcommand{\rthis}[1]{\textcolor{black}{#1}}
\begin{document}
\newcolumntype{P}[1]{>{\centering\arraybackslash}p{#1}}
\pdfoutput=1
\newcommand{\jcap}{JCAP}
\newcommand{\araa}{Annual Review of Astron. and Astrophys.}
\newcommand{\apss}{Astrophysics and Space Sciences}
\newcommand{\aj}{Astron. J. }
\newcommand{\mnras}{MNRAS}
\newcommand{\apjl}{Astrophys. J. Lett.}
\newcommand{\apjs}{Astrophys. J. Suppl. Ser.}
\newcommand{\aap}{Astron. \& Astrophys.}
\newcommand{\ssr}{Space Science Reviews}
\newcommand{\solphys}{Solar Physics}
\renewcommand{\arraystretch}{2.5}
\title{Generalized Lomb-Scargle Analysis of 22 years of Super-Kamiokande solar $^8$B neutrino data}
\author{Vibhavasu \surname{Pasumarti}}
\altaffiliation{E-mail:ep20btech11015@iith.ac.in}

\author{Shantanu \surname{Desai}}
\altaffiliation{E-mail: shntn05@gmail.com}

\begin{abstract}
We apply the generalized Lomb-Scargle periodogram to 22 years data of solar $^{8}$B neutrino fluxes detected by Super-Kamiokande. The primary motivation of this work was to check if the sinusoidal modulation at a frequency of 9.43/year (with a period of 38 days), which we had found to be marginally significant with the first five years of Super-K data, persists, with the accumulated data. We use four different metrics for the calculation of significance of any peaks in the Lomb-Scargle periodogram, which could be indicative of periodicities. We do not find any evidence for periodicity at the aforementioned frequency or any other frequency with the updated data. 
Therefore, the observed peak at 9.43/year with the first five years of Super-Kamiokande data was only a statistical fluctuation and its significance is negligible with the updated data.

\end{abstract}

\affiliation{Department of Physics, IIT Hyderabad, Kandi, Telangana-502284, India}

\maketitle
\section{Introduction}
For more than two decades, Sturrock and collaborators have argued that the $^{8}$B solar neutrino flux seen in Super-Kamiokande-I from 1996-2001~\cite{Yoo} contains a sinusoidal modulation at a frequency of 9.43/year, which corresponds to a period of around 38 days (See Ref.~\cite{Sturrock22} and references therein). They asserted that this peak is due to the synodic rotation of the solar core, for which the sidereal rotation rate is around 10.43/year~\cite{Sturrock08,Sturrock12}. The first independent analysis of the Super-K data reported $p$-values for this peak ranging from 0.007 to 0.019, depending on the analysis technique used~\cite{Sturrock05}. Another work found $p$-values from 0.018 to 0.12, depending on whether one incorporates the asymmetric errors or not~\cite{Ranucci}. However, these results were at odds with the analysis done by the Super-K collaboration, which did not find any evidence for periodicity at any frequency~\cite{Yoo}.

In order to investigate this issue, we applied the generalized Lomb-Scargle periodogram~~\cite{Lomb,Scargle,Kurster} to Super-K-I data (binned in intervals of 5-days)~\cite{Desai16}. We then calculated the $p$-value using the method proposed in ~\cite{Scargle}, as well as using Bayesian information criterion (BIC)~\cite{Krishak}. We were able to confirm the peak (previously observed in works by Sturrock and collaborators) at 9.43/year with a $p-$value of \rthis{0.015 (significance of 2.2 $\sigma$)}, having a BIC value close to 5, pointing to marginal significance using qualitative ``strength of evidence'' rules~\cite{Krishak}. This peak was also confirmed in other works~\cite{Pommesolar,Pomme2018decay}. The amplitude of this modulation at a period of 38 days is equal to 6.9\%. Most recently, it has been shown that the significance of this peak is due to the fortuitous alignment of six data points~\cite{Pomme22}. If these data points are excluded, the amplitude gets reduced to 5.3\% thereby reducing the significance of the peak. Therefore, Ref.~\cite{Pomme22} has asserted that the aforementioned peak is not real and only a statistical fluctuation. Nevertheless, the only way to resolve this imbroglio is to redo the same procedure with additional data.

Another reason for possible sinusoidal modulations in the solar neutrino flux could be due to time variations in the solar magnetic field, if the neutrino has a non-zero magnetic moment, \rthis{because of Resonant Spin Flavor precession~\cite{Pulido,Akhmedov,SK23}.} 
Finally, a periodic variation in the solar core temperature could also induce a periodicity in the solar neutrino fluxes~\cite{Ray71}.
Therefore observing such a periodic modulation could enable us to gain insights 
on a variety of physical processes in the solar interior and also some of the fundamental properties of 
neutrinos.

Twenty years after the first search for periodicities, the Super-K collaboration has recently carried out another search for periodicities with a livetime of around 22 years~\cite{SK23}. Therefore, this data would provide a ``smoking gun'' test on whether the tantalizing hints for a periodic signal at 38 days seen in the first five years of data were only a fluctuation or signatures of a real sinusoidal signal. However, no evidence for periodicity was found in this analysis. This data has also been made publicly available.

In this work, we again apply the generalized Lomb-Scargle periodogram (as our previous work~\cite{Desai16}) to the aforementioned Super-K data~\cite{SK23}. We evaluate the statistical significance using four independent methods, similar to our recent works on searching for periodicities in nuclear $\beta$-decay rates~\cite{Tejas,Dhaygude,Gautam}. This manuscript is structured as follows. We provided a brief account of the generalized Lomb-Scargle periodogram in Sect.~\ref{sec:ls}. We recap the latest Super-K search for periodicity using more than 20 years of data in Sect.~\ref{sec:skrecap}.
Our analysis and results can be found in Sect.~\ref{sec:analysis}. We conclude in Sect.~\ref{sec:conclusions}.

\section{Generalized Lomb-Scargle Periodogram}
\label{sec:ls}
The Lomb-Scargle (L-S)~\cite{Lomb,Scargle,Vanderplas,Vanderplas15,astroml} periodogram is a widely used robust technique to look for periodicities in unevenly sampled data. The main goal of the L-S periodogram is to determine the frequency ($f$) of a periodic sinusoidal signal in a time-series ($y(t)$)
as follows:
\begin{equation}
y(t)=a\cos(2\pi f t)+ b \sin(2 \pi f t).
\label{eq:yt}
\end{equation}
The L-S periodogram calculates the power as a function of frequency, from which one can assess the statistical significance for any frequency.

For this analysis, we use the generalized (or floating-mean) L-S periodogram~\cite{Kurster,Bretthorst}. The main difference with respect to the ordinary L-S periodogram is that an arbitrary offset gets added to the mean values. More details on the differences between the two implementations can be found in ~\cite{Vanderplas,Vanderplas15}.
The generalized L-S periodogram has been shown to be more sensitive than the normal one, for detecting peaks when the data sampling overestimates the mean~\cite{Vanderplas,Kurster,proc}. To determine the significance of any peak in the L-S periodogram, we must calculate its false alarm probability or $p$-value. A large number of metrics
have been constructed to estimate the $p$-value of peaks in the L-S periodogram~\cite{Scargle,Vanderplas,Pommemod,Pommesig}. We now briefly describe these myriad metrics, which we label based on the command-line options in {\tt Python}, which are used to calculate these $p$-values:
\begin{itemize}
 \item {\bf Baluev}

 This method uses extreme value statistics for stochastic process, to compute an upper-bound of the $p-$value in case of no aliasing. The analytical expression for the $p-$value using this method can be found in ~\cite{Baluev,Vanderplas}.
 
 \item {\bf Bootstrap}
 
 This method makes use of non-parametric bootstrap resampling~\cite{Vanderplas}. It applies L-S periodograms on synthetic data constructed at the same observation times as the real data. The bootstrap is the most robust estimate of the $p$-value, as it makes minimal assumptions about the periodogram distribution, and the observed times also fully account for survey window and dead-time effects~\cite{Vanderplas}. \rthis{However, the bootstrap method does not correctly account for correlated noise in the observations~\cite{astroml}. One also needs a large number of bootstrap resamples to compute the $p-$value with very good accuracy. We use 1000 bootstrap samples which can provide $p$-values with an accuracy of about 1\%. We also note that for our analysis, we are using binned data, which consists of observations averaged over a five-day period, and hence we are not fully incorporating the detector livetime and dead-time effects.}
 \item {\bf Davies}
 
 This method is similar to Baluev, but is not accurate at large $p$-values, where it shows values greater than 1~\cite{Davies}. Nevertheless, for completeness we also calculate the $p$-value using this method.
 \item {\bf Naive}
 
  This method is based on the assertion that well-separated areas in the periodogram are independent of each other.  The total number of independent frequencies is dependent on the sampling rate and observation duration~\cite{Vanderplas}. 
 \end{itemize}

\rthis{We note that all the aformentioned $p$-values are global $p$-values that evaluate the significance of peaks in the periodogram after accounting for the ``look elsewhere effect'': the trials factors associated with the fact that many frequencies are being searched at once~\cite{Gross}. However, for two specific frequencies (9.43/year and 1/year), we also calculate the local $p$-value using the expression in Table 1 of ~\cite{Baluev}. This local $p$-value has been implemented using the {\tt single} option in the L-S module provided in {\tt astropy}.}
Once the $p-$value is known, one can evaluate the significance or $Z$-score~\cite{Ganguly}. 
The smaller the $p$-value, the more significant is the peak. A rule of thumb for any peak to be deemed interesting is that $p$-value should be less than $0.05$. However for a peak to be statistically significant, its $Z-$score must be greater than 5$\sigma$~\cite{Lyons}, which corresponds to $p$-value $< 10^{-7}$.

\section{Recap of SK23}
\label{sec:skrecap}
\begin{figure}[H]
 \centering
 \includegraphics[width=\textwidth]{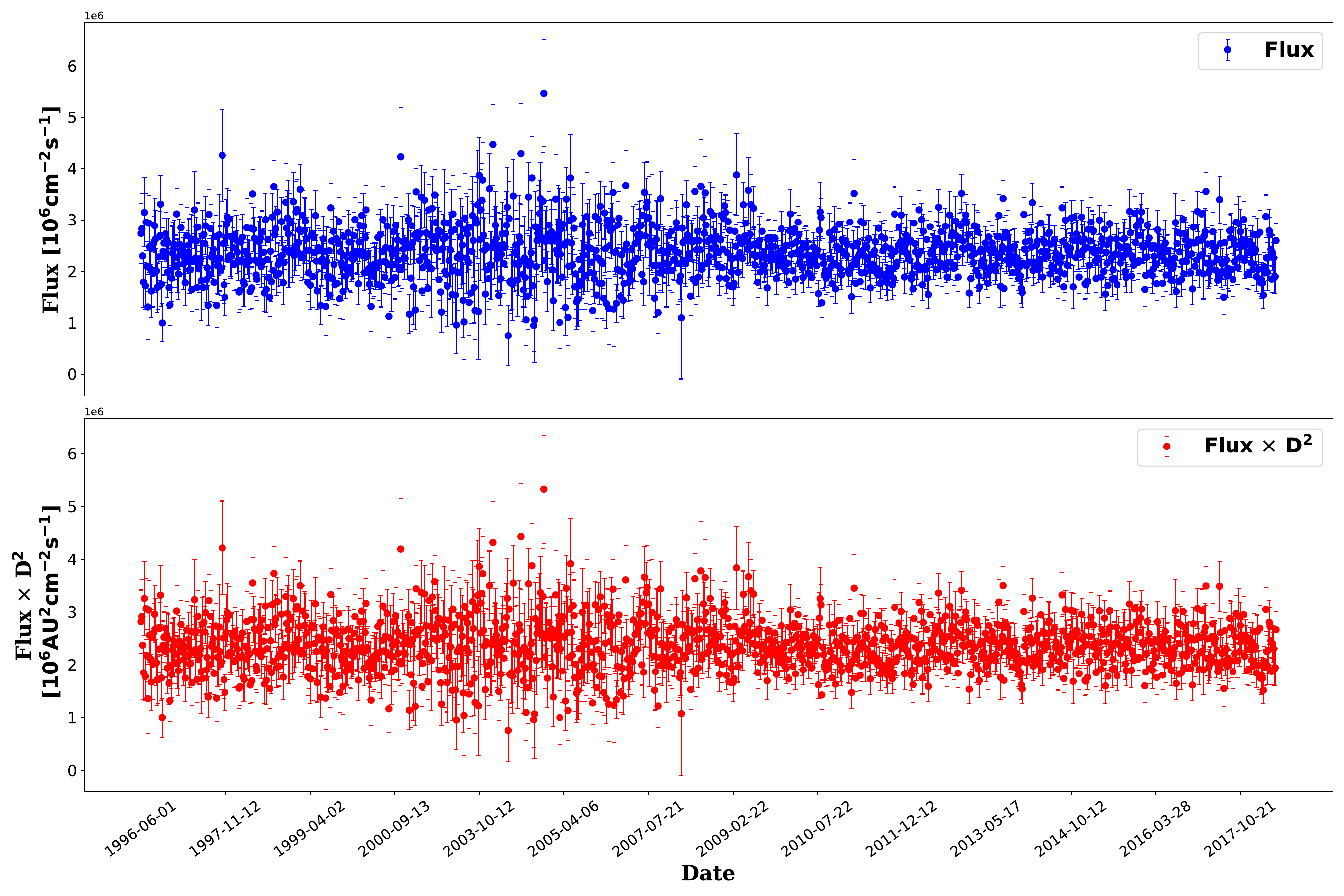}
 \caption{The Super-K $^8$B solar neutrino flux as a function of time. The top panel shows the raw flux and the bottom panel is the corrected flux rescaled by the square of the distance between the Sun and the Earth averaged over each 5-day bin. Data used for this plot has been obtained from ~\cite{SK23}. For brevity, we have ignored the X-axis bin width in the above plots.}
 \label{fig:timeseries}
\end{figure}

We briefly describe the latest search for periodicity carried out by the Super-K collaboration using more than 20 years of $^8$B solar neutrino data from 1996-2018~\cite{SK23}. Super-K is a 50 kiloton water Cherenkov detector located in the Kamioka mine in Japan which detects neutrinos with energies from MeV range~\cite{SK98} to over a TeV~\cite{SKshowering}, which has been taking data since 1996. Super-K has produced physics and astrophysics results from a wide range of topics from neutrino oscillations~\cite{Superknuosc} to dark matter~\cite{Desai04}.
The total livetime used for the analysis of solar neutrino data until 2018 is equal to 5,804 days combining four different phases of data taking spanning about 22 years. Super-K detects about 20 solar neutrino interactions per day. The total dataset was divided into 1343 time bins, where the average width of each bin is around five days.
A search for perodicities was done using the Maximum likelihood method (by incorporating the energy information in addition to the fluxes) as well as the L-S periodogram following the prescription in ~\cite{NR}. The L-S analysis was done by looking for 100,000 frequencies from $10^{-6}$/day to 0.2/day. 
No significant periodicities apart from the annual modulation due to the revolution of the Earth around the Sun were found. As a supplement to this paper, the data for 5-day interval fluxes covering the above observing period have been made publicly available. The publicly released dataset consists of mean time, start and end time of the observed data, the measured solar neutrino flux along with its upper and lower flux errors. 
However, this data does not include corrections due to the varying Earth-Sun distance. Unlike the Super-K-I dataset, these correction factors due to the eccentricity of the Earth's orbit around the Sun were not provided along with the dataset at the time of writing. So we obtained these correction factors by calculating the average Earth-Sun distance binned in 1-day intervals ($D$) for each 5-day bin. The distance to the Sun was calculated using the {\tt astropy}~\cite{astropy} module. We then scaled the raw flux and the errors by $D^2$ after normalizing the distance by 1~AU. This plot of both the uncorrected and corrected fluxes as a function of time since the start of Super-K can be found in Fig.~\ref{fig:timeseries}.

\section{Analysis and Results}
\label{sec:analysis}
We now apply a generalized L-S periodogram to this dataset. We used the L-S implementation in {\tt astropy.time\_series}~\cite{astropy} module in {\tt Python}. We provided the mid-point of each time bin, measured flux and the flux errors as inputs to the L-S periodogram. \rthis{The flux errors were obtained from an average of the asymmetric error bars. However, as a sanity check we also redid the analyses by considering the larger values among the errors for each data point, but the results are comparable to those obtained by considering the average of the errors and are not reported here.}
The recommended frequency resolution and maximum frequency up to which the generalized L-S method can robustly detect sinusoidal modulations are given by the reciprocal of five times the total duration of the dataset and five times the mean Nyquist equivalent frequency, respectively~\cite{Vanderplas}. 
For the Super-K dataset, this default frequency resolution is equal to 0.0091/year, which is used for our work. Based on the above recommendation, the maximum frequency up to which the L-S periodogram would be sensitive to any potential peaks is equal to 152.8/year. 
However for brevity, we only show the results for frequencies up to 20/year (similar to our previous work), since previous searches had shown a statistically significant result only at 9.43/year~\cite{Desai16} and the frequencies associated with solar rotation are between 8-14/year. However, we also checked if the updated data exhibited any statistically significant peaks for frequencies greater than 20/year and up to 70/year (which is equal to the width of each time bin). Since we did not find any peaks with large statistical significance above a frequency of 20/year, we do not show any results for the same.
We followed the same normalization convention for the L-S periodogram as in our previous works ~\cite{Tejas,Dhaygude,Gautam}, where the L-S power can take values between zero and one. In the appendix, we also show the results of the same analyses using only the first five years of Super-K data until July 2001.

\begin{figure}
 \centering
 \includegraphics[width=\textwidth]{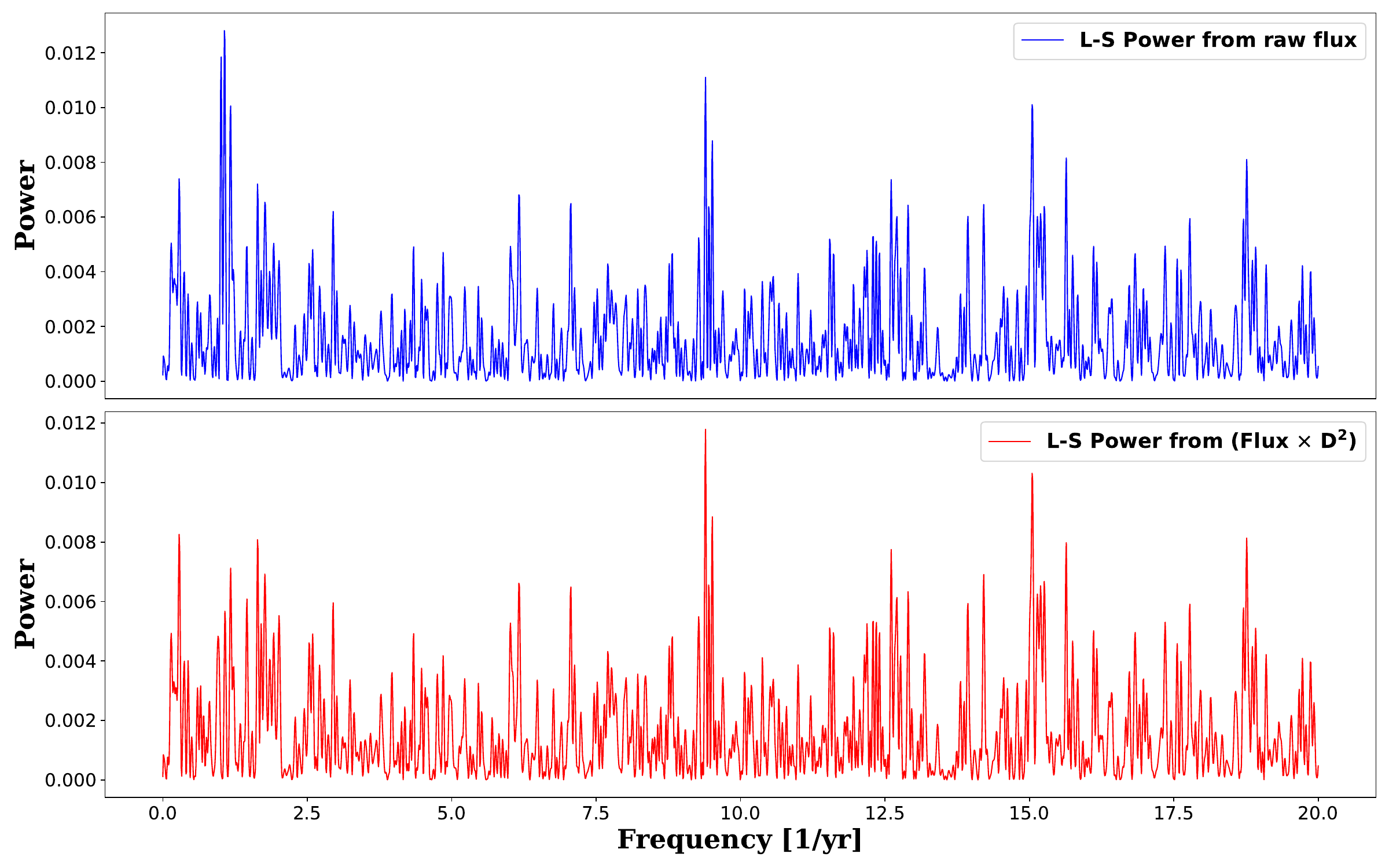}
 \caption{L-S power as a function of frequency using the generalized L-S periodogram for the raw flux (top panel) and the corrected flux (bottom panel). }
 \label{fig:LS}
\end{figure}

The L-S power as a function of frequency for both the uncorrected as well as the corrected $^8$B neutrino fluxes are found in Fig.~\ref{fig:LS}. We see a maximum peak for the uncorrected flux at a frequency of 1/year. However, this peak is not the strongest, once we rescale the fluxes by the square of the distance. We calculated the $p$-value for the first four significant peaks. These include the frequency of 9.39/year, which is close to 9.43/year previously found to be marginally significant in ~\cite{Desai16}. This table containing the L-S powers and the $p$-values (using all the four methods outlined in Sect.~\ref{sec:ls}) for the top four frequencies with the largest L-S powers can be found in Tables~\ref{tab:raw_flux} and ~\ref{tab:corr_flux}, respectively. We find that for both the corrected as well as uncorrected flux, we do not find any statistically significant peaks. All the observed $p$-values (using any method) are greater than 0.4. For the uncorrected flux, the maximum power is seen at a frequency of 1.06/year with the lowest $p$-value of 0.47 using the {\tt Naive} method. For the flux, scaled by the square of the distance, the maximum power is seen at a frequency of 9.39/year (which is \rthis{close to the frequency of 9.43/year}, which we had previously found in Super-K-I data~\cite{Desai16}). However, its statistical significance is negligible (with the minimum $p$-value of 0.71). This is in contrast to the value of 0.015 we had found for Super-K-1. Therefore, we conclude that with the latest updated data, the marginally significant peak \rthis{close to the} frequency of 9.43/year using the first five years of data has disappeared. Since no other significant peaks were found, we therefore assert that there are no periodic sinusoidal signals at any frequencies in the $^{8}$B solar neutrino flux. \rthis{
We also applied the generalized L-S periodogram to all the Super-K data post July 2001 (which was not included in the earlier analysis) and calculated the local $p-$value for the frequency of 9.39/year. We find the $p$-values at this frequency to be equal to 0.074 (1.4$\sigma$) and 0.056 (1.6$\sigma$) for the uncorrected and corrected fluxes, respectively. Therefore, the local $p-$values are not statistically significant for the frequency of 9.39/year, when analyzing the data after the phase-I of Super-K. For the frequency of 1/year, the local $p-$value is equal to 0.001  and 0.54 for the uncorrected and corrected flux, respectively using the full 22 years of Super-K data. Therefore, the uncorrected flux corresponds to a local significance of $3.1\sigma$. Once we correct for the Earth's eccentricity, the significance becomes negligible. }

To summarize, we do not find evidence for periodicities at any frequency using 22 years of Super-K data,  in accord with the results found in ~\cite{SK23}.

\begin{table}[H]
 \centering
 \begin{tabular}{|c|c|c|c|c|c|}
 \hline
   Frequency [year$^{-1}$] & L-S Power & \texttt{Bootstrap} & \texttt{Naive} & \texttt{Baluev} & \texttt{Davies} \\ \hline
   1.06 & 0.013 &   0.92 &  0.47 &   0.84 &   $>$ 1 \\
   9.39 & 0.011 &   0.99 &  0.86 &   0.99 &   $>$ 1 \\
   15.05 & 0.010 &   1.000 &  0.98 &   1.000 &  $>$ 1 \\
   18.76 & 0.008 &   1.000 &  1.000 &   1.000 &  $>$ 1 \\

\hline
   \bottomrule
 \end{tabular}
 \caption{L-S power and $p$-value (last four columns) for the uncorrected $^{8}$B Super-K flux using four different methods for the four frequencies showing the largest L-S powers in descending order. All the $p$-values are greater than 0.5, implying that none of the peaks are statistically significant, and there is no evidence for sinusoidal modulations in the data.}
 \label{tab:raw_flux}
\end{table}

\begin{table}[h]
 \centering
 \begin{tabular}{|c|c|c|c|c|c|}
  \hline
   Frequency [year$^{-1}$] & L-S Power & \texttt{Bootstrap} & \texttt{Naive} & \texttt{Baluev} & \texttt{Davies} \\ \hline
   9.39 & 0.012 &   0.98 &  0.71 &   0.971 &   $>$ 1 \\
   15.05 & 0.010 &   1.000 &  0.96 &   1.000 &   $>$ 1 \\
   0.28 & 0.008 &   1.000 &  1.000 &   1.000 &  $>$ 1 \\
   18.76 & 0.008 &   1.000 &  1.000 &   1.000 &  $>$ 1 \\
   
   \hline
 \end{tabular}
 \caption{L-S power and $p$-value (last four columns) for the rescaled $^{8}$B Super-K flux after correcting for the eccentricity of the Earth's orbit, using four different methods for the four frequencies showing the largest L-S powers in descending order. Similar to the uncorrected flux, all the $p$-values are greater than 0.5, implying that none of the peaks are statistically significant, and there is no evidence for sinusoidal modulations in the data.}
 \label{tab:corr_flux}
\end{table}

\section{Conclusions}
Multiple groups have found evidence for sinusoidal modulations in the $^8$B solar neutrino flux observed using the first five years of Super-K data at a frequency of around 9.43/year, corresponding to a period of 38 days. These peaks were asserted to be due to the synodic rotation of the solar core.
Our own analysis (in 2016) of this data using the generalized L-S periodogram found a $p$-value of around 0.015. In November 2023, the Super-K collaboration updated its results for periodicity searches using 22 years of data using two independent methods, one of which includes the (normal) L-S analysis. This work did not find evidence for any statistically significant peak (apart from the variation due to Earth's orbit around the Sun)~\cite{SK23}. The dataset used for this analysis was also made publicly available.

We carried out an independent search for periodicity with the same data using the generalized L-S periodogram to ascertain if the updated data again contains a modulation at \rthis{or near} 9.43/year, and if its detection significance is enhanced with five times more data. We analyzed both the raw solar neutrino fluxes and also the fluxes rescaled by the square of the distance between the Earth and the Sun. We estimated the $p$-values using four independent methods. Our plots for the generalized L-S power as a function of frequency can be found in Fig.~\ref{fig:LS}. The $p$-values we found for the top four frequencies in descending order of their L-S powers can be found in Table~\ref{tab:raw_flux} and Table~\ref{tab:corr_flux}, respectively.
We no longer find any statistically significant peak at a frequency \rthis{close to} 9.43/year. For the corrected flux, the minimum $p$-value we found was equal to 0.71 (at a frequency of 9.39/year), which is not significant. Therefore, we conclude that the entire 22 years of Super-K data no longer contains any signatures of the sinusoidal modulations \rthis{close to}  a frequency of 9.43/year \rthis{(period of 38 days)} or any other frequency. Our results are also in accord with the corresponding analyses carried out by the Super-K Collaboration.

In the spirit of open science, we have made our analysis codes for this work publicly available, which can be found at \url{https://github.com/DarkWake9/Project-LP}.

\label{sec:conclusions}
\section*{Acknowledgements}
We are grateful to the Super-K collaboration for making their solar neutrino dataset publicly available, to Jonghee Yoo for clarifications about the released dataset \rthis{and to the anonymous referees for very useful feedback on the manuscript.}
\bibliography{main}
\section*{APPENDIX}
\begin{figure}
  \centering
  \includegraphics[width=\textwidth]{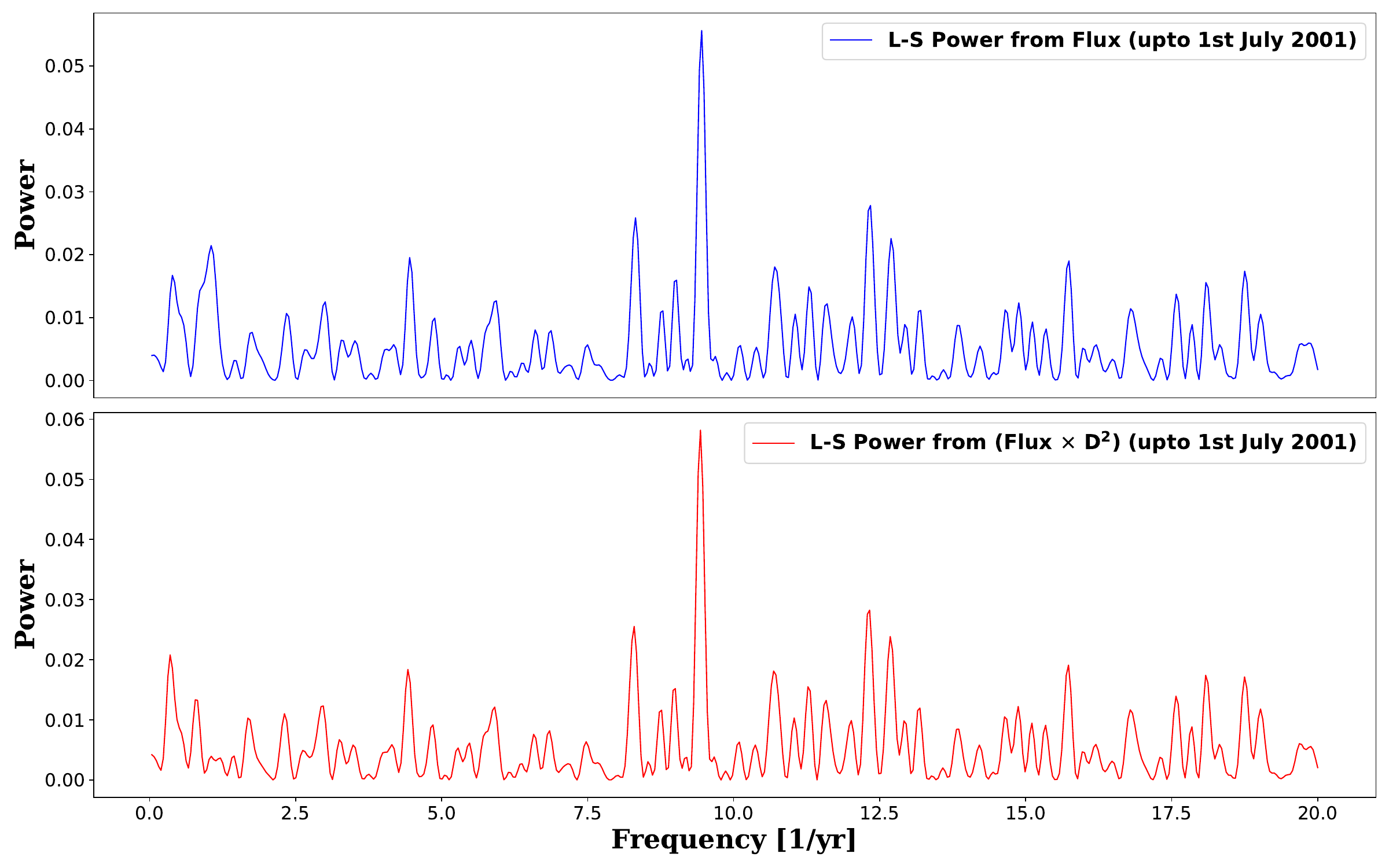}
  \caption{L-S Periodogram of Super-K data from April 1996 to July 2001}
  \label{fig:Jul2001}
\end{figure}

\rthis{In order to provide a direct comparison with the results using data only from the phase-I of Super-K, we redo the same analysis as done in the main paper using data only up to July 2001. For this purpose, we use the same data provided in ~\cite{SK23}, but culled all the data after July 2001, and used the same correction factors to correct for Earth's orbit around the Sun, which we had calculated. Our results for the same can be found in Fig~\ref{fig:Jul2001}, both with and without the correction factors. Note that we had analyzed the same data (based on the data and correcton factors provided in ~\cite{Yoo}) in ~\cite{Desai16} using the generalized LS implemenation in the {\tt astroML} library.) Once again, the highest peak is seen at a frequency of 9.43/year. The $p-$values using all the four methods are shown in Table~\ref{tab:corr_flux2001}. The minimum $p-$ value is found for the {\tt Naive} method with a $p$-value of 0.028 (after correcting for the distance). This corresponds to a $Z$-score of 1.9$\sigma$. This is close to (but slightly larger than) the value of 0.015 we had found in ~\cite{Desai16}. }

\begin{table}[h]
 \centering
 \begin{tabular}{|c|c|c|c|c|c|c|}
  \hline
   & Frequency [year$^{-1}$] & L-S Power & \texttt{Bootstrap} & \texttt{Naive} & \texttt{Baluev} & \texttt{Davies} \\ \hline
   Not Corrected Flux & 9.43 & 0.056 & 0.17 &   0.045 &  0.14 &   0.15\\
   Corrected Flux & 9.43 & 0.058 & 0.107 &   0.028 &  0.091 &   0.096\\   
   \hline
 \end{tabular}
 \caption{L-S power and $p$-value (last four columns) for $^{8}$B Super-K flux from April 1996 to July 2001 using four different methods for the four frequencies showing the largest L-S powers. }
 \label{tab:corr_flux2001}
\end{table}

\end{document}